\UseRawInputEncoding 
\documentclass[reprint,superscriptaddress,amsmath,amssymb,
aps,prl,floatfix]{revtex4-1}

\usepackage{bm}
\usepackage{float}
\usepackage{graphicx}
\usepackage[usenames,dvipsnames]{color}
\usepackage[normalem]{ulem}
\usepackage[svgnames]{xcolor}
\usepackage{bm}
\usepackage{multirow}
\usepackage{float}
\usepackage{titlesec}

\begin{document}
\preprint{APS/123-QED}

\title{Mesoscopic tunneling in strontium titanate}

\author{Benoît Fauqu\'e}%
 \email{benoit.fauque@espci.fr}
\affiliation{JEIP,  USR 3573 CNRS, Coll\`ege de France, PSL Research University, 11, place Marcelin Berthelot, 75231 Paris Cedex 05, France}

\author{Philippe Bourges}
\affiliation{Laboratoire L\'eon Brillouin, CEA-CNRS, Universit\'e Paris-Saclay, CEA Saclay, 91191 Gif-sur-Yvette, France}

\author{Alaska Subedi}
\affiliation{CPHT, CNRS, Ecole Polytechnique, IP Paris, F-91128 Palaiseau, France}

\author{Kamran Behnia}
\affiliation{Laboratoire de Physique et d'Étude des Matériaux (ESPCI Paris - CNRS - Sorbonne Universit\'e), PSL Research University, 75005 Paris, France}

\author{Benoît Baptiste}
\affiliation{IMPMC-Sorbonne Université and CNRS, 4, place Jussieu, 75005 Paris, France}

\author{Bertrand Roessli}
\affiliation{Laboratory for Neutron Scattering and Imaging, Paul Scherrer Institut,
Villigen, Switzerland}

\author{Tom Fennell}
\affiliation{Laboratory for Neutron Scattering and Imaging, Paul Scherrer Institut,
Villigen, Switzerland}

\author{St\'ephane Raymond}
\affiliation{Univ. Grenoble Alpes, CEA, IRIG, MEM, MDN, 38000 Grenoble, France}

\author{Paul Steffens}
\affiliation{Institut Laue-Langevin, 71 Avenue des Martyrs, 38042 Grenoble Cedex 9, France}

\date{\today}

\begin{abstract}
Spatial correlation between atoms can generate a depletion in the energy dispersion of acoustic phonons. Two well known examples are rotons in superfluid helium  and  the Kohn anomaly in metals. Here we report on the observation of a large softening of the transverse acoustic mode in quantum paraelectric SrTiO$_3$ by means of inelastic neutron scattering. In contrast to other known cases,  this softening occurs at a tiny wave vector implying spatial correlation extending over a distance as long as 40 lattice parameters. We attribute this to the formation of mesoscopic fluctuating domains due to the coupling between local strain and quantum ferroelectric fluctuations. Thus, a hallmark of the ground state of insulating SrTiO$_3$ is the emergence of hybridized optical-acoustic phonons. Mesoscopic fluctuating domains play a role in quantum tunneling, which impedes the emergence of a finite macroscopic polarisation.

\end{abstract}

\maketitle

In solids and fluids the energy cost of an elastic modulation, $\hbar \omega$, is set by its wave-vector $q$ ($=\frac{2\pi}{\lambda}$ where $\lambda$ is its wavelength). This relation is encoded in the phonon dispersion $\omega(q)$. For long wavelengths, this energy is low and the phonon dispersion is linear in $q$ and the slope defines the sound velocity. When $\lambda$ becomes comparable with the interatomic distance, the discreteness of the lattice in solids gives rise to saturation of $\omega$ and a standing wave with a zero group velocity. A few exceptions to this monotonic dispersion are known, such as superfluid $^4$He \cite{Henshaw1961} and metals displaying a Kohn anomaly in their phonon spectrum \cite{Kohn1959,Renker1973}. In the case of helium, a phonon branch dubbed roton displays a minimum at finite $q$. Roton-like dispersion has been recently identified in other quantum fluids \cite{Godfrin2012,Chomaz2018}, and has been predicted to occur in chiral materials \cite{Kishine2020} or classical acoustic metamaterials \cite{Chen2021}. Broadly speaking, the presence of a soft acoustic branch indicates a tendency towards local order driven either by correlations between atoms and/or to the proximity of an instability in the system.

In this study, we report on the observation of a large softening of the transverse acoustic (TA) mode by means of inelastic neutron scattering in  SrTiO$_3$, a quantum paraelectric solid \cite{Muller1979}. In contrast with other cases of roton-like dispersion, the softening occurs at a very small $q\approx$ 4 $\times 10^{-2}$\AA$^{-1}$, pointing to the existence of mesoscopic fluctuating domains as wide as $\approx$ 16 nm.  

Ferroelectric ordering at low temperature is aborted in SrTiO$_3$  by zero-point quantum fluctuations \cite{Schneider1976,Shin2021,esswein2021}. The large dielectric constant in this state has made SrTiO$_3$ one of the most widely used substrates for growing oxide heterostructures \cite{Zubko2011}. The bulk solid has recently become a subject of renewed fundamental interest \cite{CollignonRev2019}. The temperature dependence of the dielectric constant has been associated with quantum criticality \cite{Rowley2014,Coak12707,Chandra_2017}. The dilute superconducting ground state  which emerges upon doping \cite{Lin2013} and its interplay with ferroelectricity \cite{Rischau2017,Ahadi2019} has attracted much attention, as well as the electronic transport properties of the dilute metal \cite{Lin2015sc,Edge2015,Marel2019,Collignon2020,Collignon2021,Kumar2021,kiseliov2021theory}. The phonon softening observed here introduces a novel ingredient to this picture and  may play a key role in what sets the ground state of SrTiO$_3$  \cite{Muller1991,Lemanov2002,Coak12707} apart. We track its origin to the coupling between a transverse acoustic branch and quantum ferroelectric fluctuations driven by the strong anharmonic lattice dynamics. 

\begin{figure}
\begin{center}
\centering
\includegraphics[width=0.49\textwidth]{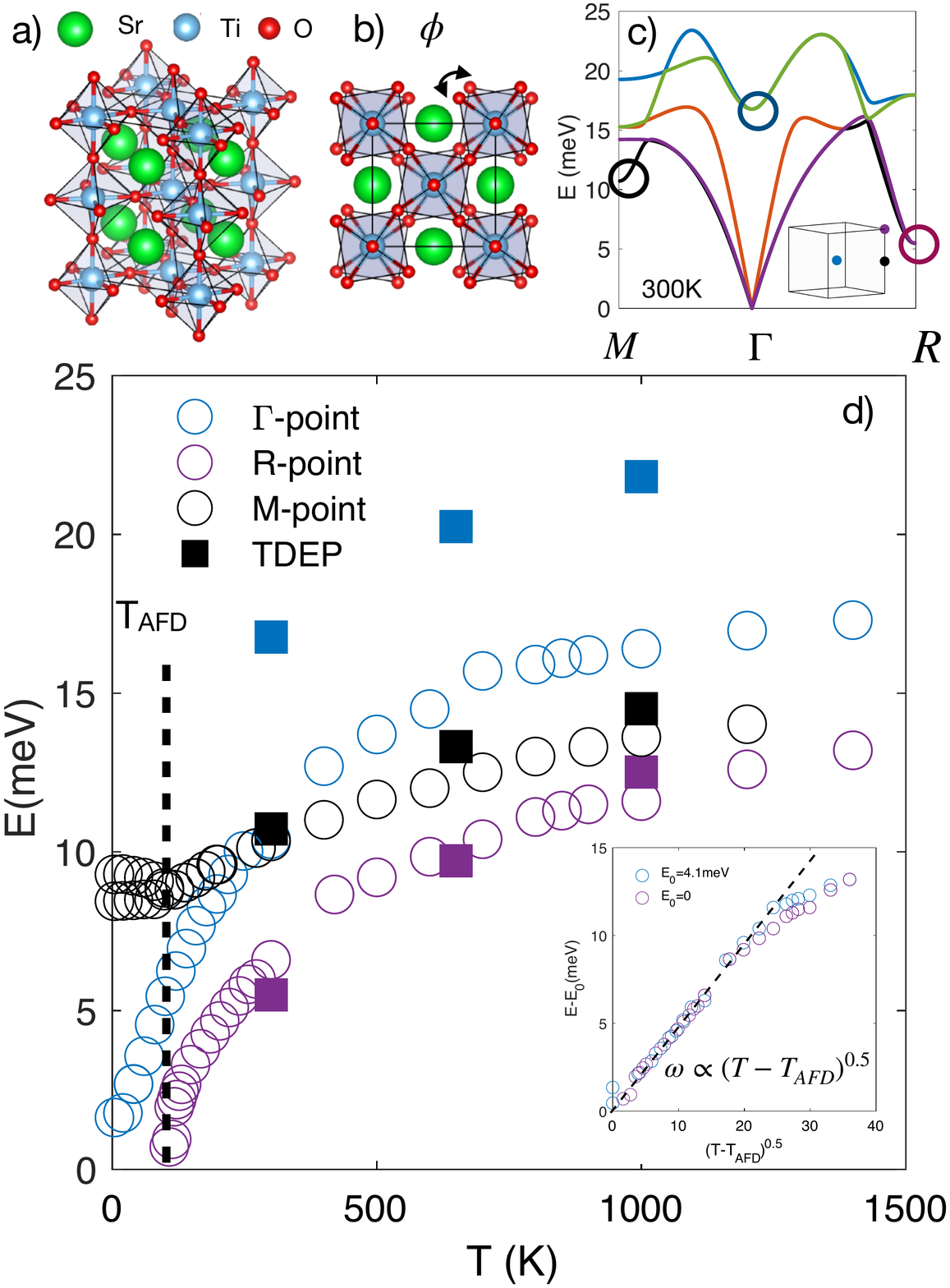}
\caption{{\bf{Soft phonons modes in SrTiO$_3$:} a) Lattice structure of SrTiO$_3$  the titanium atoms are in the center of oxygen polyhedras. Atoms are sketched according to their atomic radii. b) View along the $c$ axis in the antiferrodistortive (AFD) phase below T$_{AFD}$ = 105 K. c) Phonon spectrum in the cubic phase of SrTiO$_3$ along the $M$--$\Gamma$--$R$ direction at T = 300 K according to the temperature-dependent effective potential (TDEP) method (see \cite{SM}). Inset: Brillouin zone of the cubic phase of SrTiO$_3$ with the three high symmetry points $\Gamma$ (in blue), $R$ (in purple) and $M$ in black. d) Temperature dependence of the soft modes at the $\Gamma$ ($\bf{Q}$ = (2,0,0) or equivalent), $R$ (1.5,0.5,0.5) and $M$ points (1.5,0.5,0). Open circle symbols are the experimental data (see \cite{SM} for the energy scans). Square closed symbols are the results found by the TDEP method. Insert: energy curves of the ${R}$ mode and that of the ${\Gamma}$ mode shifted down by 4.1 meV, and both plotted as function of $(T-T_{AFD})^{0.5}$ where $T_{AFD}$ =105 K.}}
\label{Fig1}
\end{center}
\end{figure}

Like other perovskites, SrTiO$_3$ exhibits a cubic structure at high temperature [shown in Fig.~\ref{Fig1}(a)] that is distorted to a phase with a lower symmetry phase upon cooling. Two transverse phonon modes soften during this transition. The first one, located at the $R$ point of the Brillouin zone, attains zero frequency at T$_{AFD}$ = 105 K. This is the well-studied antiferrodistortive (AFD) transition where two neighboring oxygen octahedra rotate in opposite directions, as shown on Fig.~\ref{Fig1}(b). The second, a transverse optical mode located at the $\Gamma$ point (TO$_{\Gamma}$), saturates at $\approx2$ meV at low temperature and gives rise to the quantum paraelectric behavior. Both modes have been separately investigated by inelastic neutron scattering \cite{Shirane1969, Yamada1969}, Raman \cite{Fleury1968, Vogt95}, and optical spectroscopy \cite{Sirenko2000, vanMechelen2010}. We performed a systematic study of the phonon modes using inelastic neutron scattering at the $\Gamma$, $R$ and $M$ points over a very large temperature range 4--1400 K. The obtained energy scans (shown in \cite{SM}) have been fitted with a damped harmonic oscillator (DHO). The energies of the fitted modes are reported in Fig.~1(d). The soft modes at $\Gamma$ and $R$ have the same temperature dependence. In contrast the ${M}$ mode softens weakly and saturates at $\sim$9 meV. When shifted by 4 meV, the energy of the $\Gamma$ mode ($\omega_{\Gamma}$) perfectly overlaps with the energy of the $R$ mode ($\omega_{R}$) on the whole temperature range [see insert of Fig.~\ref{Fig1}(d)]. Given that the two modes belong to two distinct phonon branches [see Fig.~\ref{Fig1}(c)], this is remarkable. The two modes soften simultaneously from 1400 K following a mean-field type temperature dependence \cite{Yamada1969,Cowley1996} with the same prefactor. This observation implies that there is no primary mode driving the other one.

\begin{figure}
\begin{center}
\includegraphics[angle=0,width=8.5cm]{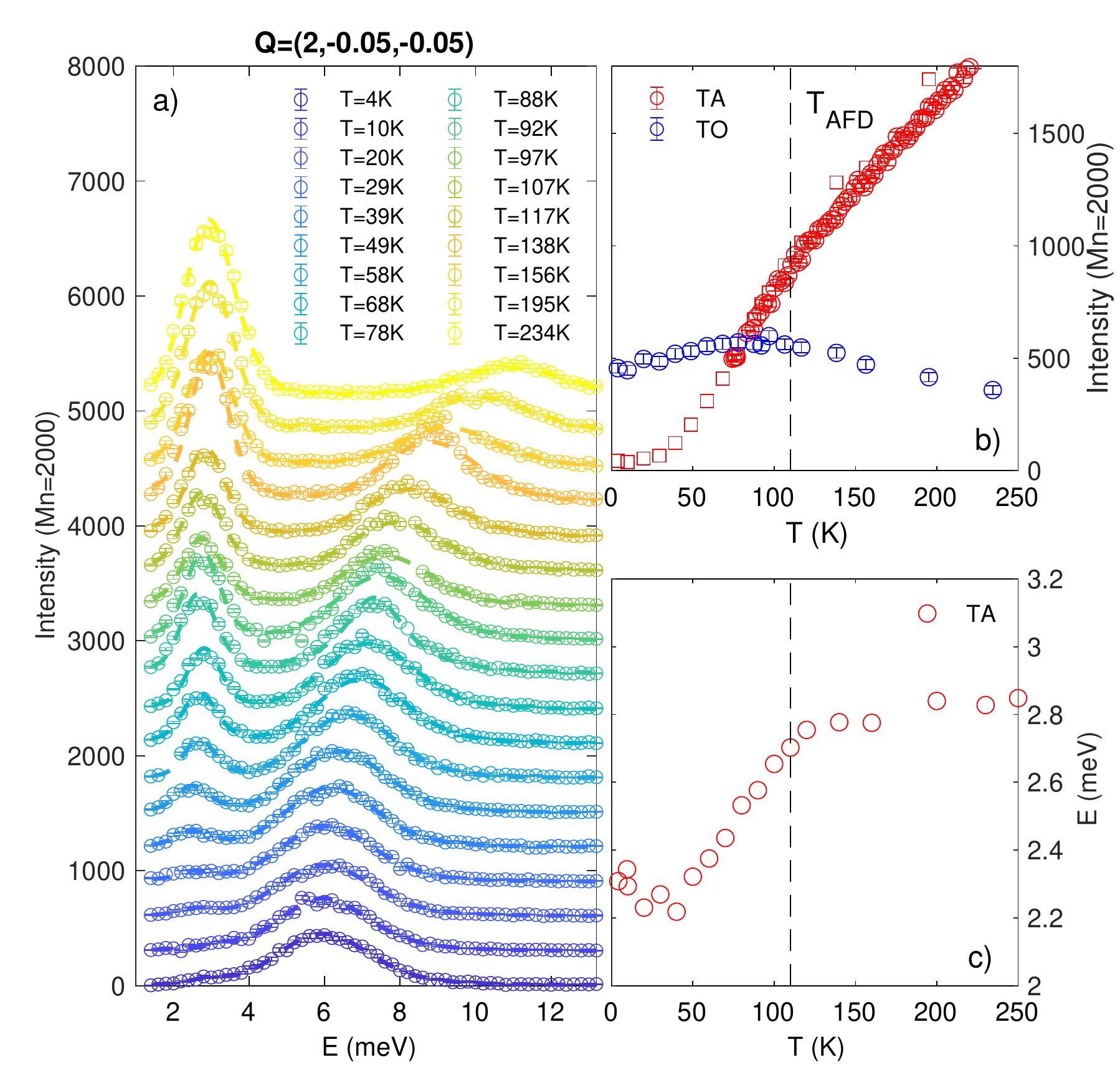}
\caption{{\bf{TA-TO coupling in the quantum paraelectric phase of SrTiO$_3$: a) Energy scans at $\bf{Q}$ = (2,-0.05,-0.05) for T = 250 K down to 4 K. Fits including a convolution with the experimental resolution are shown in dotted lines (see \cite{SM}). b) Intensity of the acoustic (in red) and optical modes (in blue) as a function of temperature. c) energy position of the acoustic mode deduced from a). We note a loss of intensity of the TA mode below $T_{AFD}$ accompanied by a softening of the TA branch }}}.
\label{FigTemp}
\end{center}
\end{figure}

\begin{figure*}
\begin{center}
\makebox{\includegraphics[width=1\textwidth]{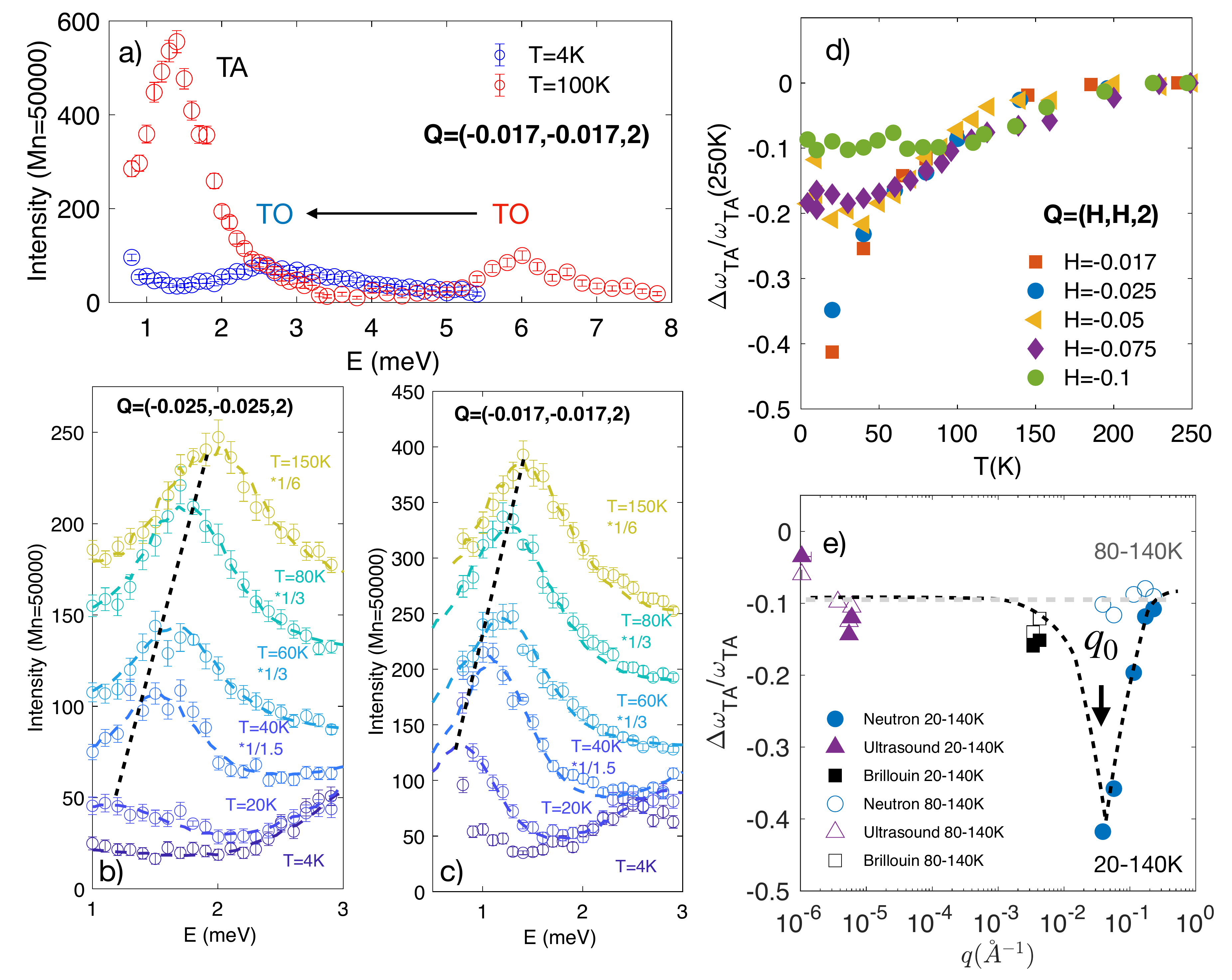}}
\caption{ {\bf{Softening of the TA branch }: a) Energy scans at $\bf{Q}$  = (-0.017,-0.017,2) for $T$ = 4 K (in blue) and 100 K (in red). Low energy scans at b) $\bf{Q}$ = (-0.025,-0.025,2) and c) $\bf{Q}$ = (-0.017,-0.017,2) from 150 K down to 4 K. Curves are shifted and renormalized to underline the large softening shift of the TA branch. Fits with DHO including a convolution with the experimental resolution are shown in dotted lines (see \cite{SM}). d) Normalized temperature dependence of the TA mode at different $\bf{Q}$ vectors. e) Comparison of the magnitude of the softening of the TA branch with three different probes between 20  and 140 K: neutrons (this work), ultra-sound \cite{Balashova1996,Rehwald1970,Rehwalda1970,Scott1997,Luthi1970}  and Brillouin \cite{Laubereau1970,Hehlen1996}. Note the horizontal logarithmic scale.}}
\label{Fig3}
\end{center}
\end{figure*}

This experimental observation of the softening of the two phonon modes at $\Gamma$ and $R$ points contrasts with what is expected by density functional theory calculations based on the harmonic approximation. They predict unstable modes not only at the $\Gamma$ and $R$ points but also at the $M$ point \cite{Lasota1997}. This discrepancy between theory and experiment can be removed by including anharmonic effects in theory. This was done with a temperature-dependent effective potential (TDEP) method, recently developed and applied to the phonon spectrum of SrTiO$_3$ \cite{Tadano2015,Zhou2018,Fumega2020,Delaire2020,vanRoekeghem2021}, which renormalizes the energies and the eigenvectors of the phonon spectrum. Square symbols in Fig.~\ref{Fig1}(d) show the energy position of these three modes found by our calculations at $T$ = 300, 650 and 1000 K (see \cite{SM} for more details). There is an excellent agreement for the $R$ and $M$ points and the relative temperature dependence of all three modes. At the $\Gamma$ point the theoretical energy position overestimates  $\omega_{\Gamma}$, like in previous calculations \cite{Tadano2015,Zhou2018,Fumega2020,Delaire2020}, suggesting that effects beyond phonon anharmonicity play a role in accurately describing this mode.

\begin{figure*}[]
\begin{center}
\makebox{\includegraphics[width=1\textwidth]{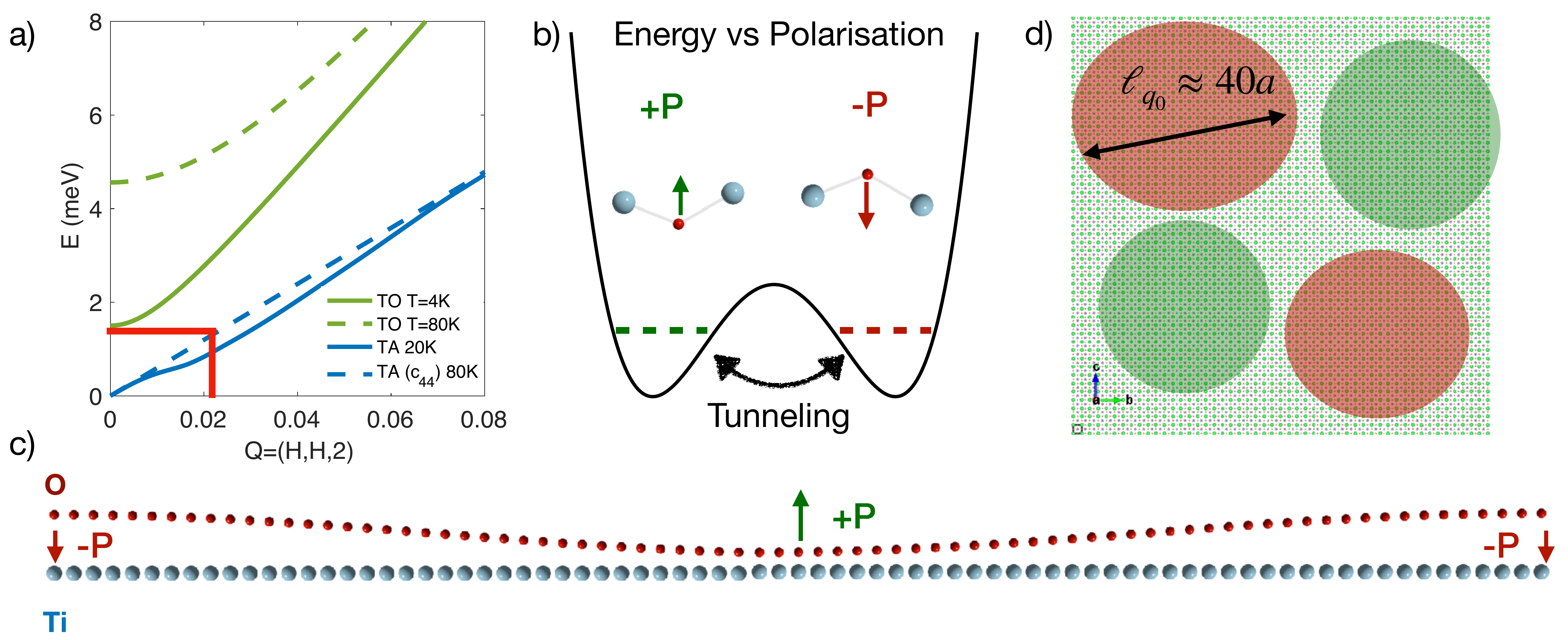}}
\caption{ {\bf{ Ground state of the quantum paraelectric SrTiO$_3$: a) Dispersion of the TO (green) and TA (blue) modes at high (dot lines) and low (full-lines) temperature. The TA dispersion at T = 80 K assumes a constant sound velocity, $v_{44}$ = 4000 m.s$^{-1}$ \cite{Rehwald1970}. The TA dispersion at T = 20 K is deduced from the softening (dotted line) shown on Fig.~\ref{Fig3}(e)). The red lines indicate the position in $Q$ where the largest coupling between the TO and TA branches is expected when $\omega_0 \approx v_{44}q_0$. b) Sketch of the free energy: the long-range polarized order is not stabilised due to tunneling processes between the lowest energy levels (dashed red and green lines) of the double well potential. c) Sketch of a transverse elastic deformation over a length scale of $\ell_0=\frac{2\pi}{q_0}$ as wide as 40 unit cells along  (1,0,0).  d) This leads to fluctuating polarized domains of mesoscopic size of both signs.}}} 
\label{Fig4}
\end{center}
\end{figure*}

A second manifestation of the anharmonicity at work in SrTiO$_3$ occurs at low-temperature. Two peaks are visible in the energy scan at $\bf{Q}$ = $(2,-0.05,-0.05)$ [see Fig.~\ref{FigTemp}(a)]. They correspond to the TA (corresponding to the elastic modulus $c_{44}$) and the TO$_{\Gamma}$ modes. As the temperature decreases, the softening of the TO$_{\Gamma}$ is accompanied by a dramatic loss of intensity of the acoustic branch. This effect, discovered more than fifty years ago \cite{Yamada1969}, is a consequence of hybridization between the TO$_{\Gamma}$ and TA modes \cite{Axe1970}. In Landau's framework of phase transitions, this is the result of coupling between strain and the gradient of fluctuations in the electric polarization \cite{Hehlen1998}. Only recently, it has been explained by a first-principles approach invoking anharmonic coupling between the renormalized phonons \cite{Delaire2020}. 

Our systematic study of the temperature and $\bf{Q}$ dependence  (shown in Figs.~\ref{FigTemp} and \ref{Fig3}) reveals two new facts. First, the loss of intensity is enhanced below $T_{AFD}$ [see Fig.~\ref{FigTemp}(b)]. Second, and most importantly, it is accompanied by a large phonon softening of the TA mode at a finite low $Q$-vector. Measurements at lower-$\bf{Q}$ [see Figs.~\ref{Fig3}(a--c)] show that the loss of intensity affects the whole TA branch. However, softening is largest at lower-$\bf{Q}$. Fig.~\ref{Fig3}(d) shows the temperature dependence of the normalised amplitude of the softening, $\frac{\Delta\omega_{TA}}{\omega_{TA}(250K)}=\frac{\omega_{TA}(T)-\omega_{TA}(250K)}{\omega_{TA}(250K)}$ at $\bf{Q}$ = ($H$,$H$,2) for $H$=$-0.017$ to $-0.1$. At $H$ = $-0.1$ and $T$= 20 K, the softening was found to be about 10$\%$, in agreement with previous measurements \cite{Yamada1969,Courtens1993,Hehlen1996,Delaire2020}. As $H$ goes to zero, $\frac{\Delta\omega_{TA}}{\omega_{TA}(250K)}$ increases up to 0.4 at $H$ = $-0.017$, the lowest ${\bf{Q}}$ vector where the TA branch could be measured in this study, see \cite{SM}.

Interestingly, this result contrasts with the elastic constant measured at lower energies by ultra-sound technique (in the range of 10-100 MHz) \cite{Balashova1996,Rehwald1970,Rehwalda1970,Scott1997,Luthi1970}  and Brillouin spectroscopy (in the range of 20 GHz) \cite{Laubereau1970,Hehlen1996,Carpenter2007}. Fig.~\ref{Fig3}(e) shows the magnitude of the softening between 20 K and 140 K (closed symbols) and between 80K and 140K (open symbols) as a function of the $q$ values (= $\omega/v_{44}$ where $v_{44}$ is the sound velocity of the c$_{44}$ mode) probed using those three techniques.  Both ultra-sound and Brillouin measurements show a sharp drop of  $v_{44}$ of 10$\%$ at T$_{AFD}$. Below T$_{AFD}$, a smooth and weak variation of about a few per cent has been detected, in contrast with the 40$\%$ drop observed by neutron scattering. 

Fig.~\ref{Fig3}(e) implies that the softening of the TA mode upon cooling is most prominent at a finite q$_{0}$ $\approx$ 0.04 ${\rm \AA}^{-1}$ below T$_{AFD}$ (see the dashed lines linking the three sets of low-temperature data). Note that one can not exclude that the softening occurs at a $q$-vector below our lowest data point and above what is resolved by Brillouin scattering, meaning that $\ell_0=\frac{2\pi}{q_0}$ could be larger. In any case, the data implies the existence of spatial correlation on a length scale of at least $\ell_{q_0} \simeq$ 40$a \approx 16$ nm where $a=$ 0.39 nm is the lattice parameter. The softening of the TO$_{\Gamma}$ is therefore accompanied by structural fluctuations extending over mesoscopic distances. We note that a similar softening can be found in KTaO$_3$ \cite{Axe1970} around q$_{0}$ $\approx$ 0.2 ${\rm \AA}^{-1}$ pointing to domain size five times shorter than in SrTiO$_3$. 

Are these fluctuating domains intrinsic? In an extrinsic scenario, the length scale would require a defect density of 2 $\times$ 10$^{17}$ cm$^{-3}$ in our undoped SrTO$_3$ samples. Such a large concentration is unlikely since, with controlled doping, one can attain a carrier concentration much less than this and detect their corresponding quantum oscillations \cite{Lin2013}. Therefore, our observation calls for an intrinsic mechanism invoking coupling between two transverse modes. Such a scenario was put forward as early as 2007 by Bussmann-Holder and co-workers \cite{Bussmann-Holder2007}, who found that coupling between the TO$_{\Gamma}$ and TA mode generates dynamical polarizability and ferroelastic clusters. However, the length scale calculated down to $\approx$ 40 K is ten times shorter than the one observed by our experiment.

The coupling between the TA and TO$_{\Gamma}$ modes is the largest when both mode are the closest, \textit{i.e.} $\omega_{TO_\Gamma} \approx v_{44}q$. With a $v_{44}$ = 4500 $m.s^{-1}$ \cite{Rehwald1970}, it occurs at $H$=0.02 or $q$= 0.05 \AA$^{-1}$ where the largest softening is indeed found [see the red line in Fig.~\ref{Fig4}(a)]. The remarkably large domain size is a consequence of the low energy value of the soft ferroelectric mode.
 
Up to now, the quantum paraelectric regime of SrTiO$_3$ was believed to be only driven by the soft TO$_{\Gamma}$ mode \cite{Schneider1976, Muller1979,Shin2021,esswein2021}. In this picture, the ferroelectric order is aborted due to quantum tunneling between two zero-point energy levels associated with two adjacent titanium-oxygen bonds [see Fig.~\ref{Fig4}(b)]. Our result reveals a missing ingredient to this picture. The ground state of SrTiO$_3$ is formed by an hybridised optical-acoustic phonon modes where quantum fluctuations are accompanied by fluctuating domains of mesoscopic length. The vanishing of average polarisation occurs thus in the context of quantum tunneling processes involving about 10$^5$ ($\approx$40$^3$) atomic unit cells acting cooperatively as sketched in  Fig.~\ref{Fig4}(c). This state can be viewed as an instanton liquid, where thanks to dynamic clusters, despite zero average polarization notwithstanding, the mean square of polarization remains finite and may even increase with decreasing temperature \cite{Ktitorov1998}.

Hybridization between optical and acoustic phonons reported here appears essential for understanding the unusual thermal conductivity \cite{Martelli2018}, the sizeable thermal Hall effect \cite{Xiaokang2020}, the temperature dependence of the electric permittivity \cite{Coak12707, Rowley2014} and the dynamics of structural domains \cite{Kustov2020} of SrTiO$_3$. Our results would also impact the ongoing discussion on the origin of the superconductivity in lightly-doped SrTiO$_3$. Coupling between electrons and the TO$_{\Gamma}$ mode plays a key role in a number of theoretical scenarios \cite{Edge2015,Marel2019,kiseliov2021theory}. We have shown that a hybridized TO$_{\Gamma}$-TA phonon mode is to be taken into account.

\begin{acknowledgments}
We thank A. Bussmann-Holder, B. Hehlen, D. Maslov, I. Paul, E. Salje and T. Weber for useful discussions.  This work was supported by JEIP-Coll\`{e}ge de France, by the Agence Nationale de la Recherche (ANR-18-CE92-0020-01; ANR-19-CE30-0014-04), GENCI (grants A0090911099, A0110913028).  
\end{acknowledgments}


\bibliography{STOInelastic}

\begin{thebibliography}{60}%
\makeatletter
\providecommand \@ifxundefined [1]{%
 \@ifx{#1\undefined}
}%
\providecommand \@ifnum [1]{%
 \ifnum #1\expandafter \@firstoftwo
 \else \expandafter \@secondoftwo
 \fi
}%
\providecommand \@ifx [1]{%
 \ifx #1\expandafter \@firstoftwo
 \else \expandafter \@secondoftwo
 \fi
}%
\providecommand \natexlab [1]{#1}%
\providecommand \enquote  [1]{``#1''}%
\providecommand \bibnamefont  [1]{#1}%
\providecommand \bibfnamefont [1]{#1}%
\providecommand \citenamefont [1]{#1}%
\providecommand \href@noop [0]{\@secondoftwo}%
\providecommand \href [0]{\begingroup \@sanitize@url \@href}%
\providecommand \@href[1]{\@@startlink{#1}\@@href}%
\providecommand \@@href[1]{\endgroup#1\@@endlink}%
\providecommand \@sanitize@url [0]{\catcode `\\12\catcode `\$12\catcode
  `\&12\catcode `\#12\catcode `\^12\catcode `\_12\catcode `\%12\relax}%
\providecommand \@@startlink[1]{}%
\providecommand \@@endlink[0]{}%
\providecommand \url  [0]{\begingroup\@sanitize@url \@url }%
\providecommand \@url [1]{\endgroup\@href {#1}{\urlprefix }}%
\providecommand \urlprefix  [0]{URL }%
\providecommand \Eprint [0]{\href }%
\providecommand \doibase [0]{http://dx.doi.org/}%
\providecommand \selectlanguage [0]{\@gobble}%
\providecommand \bibinfo  [0]{\@secondoftwo}%
\providecommand \bibfield  [0]{\@secondoftwo}%
\providecommand \translation [1]{[#1]}%
\providecommand \BibitemOpen [0]{}%
\providecommand \bibitemStop [0]{}%
\providecommand \bibitemNoStop [0]{.\EOS\space}%
\providecommand \EOS [0]{\spacefactor3000\relax}%
\providecommand \BibitemShut  [1]{\csname bibitem#1\endcsname}%
\let\auto@bib@innerbib\@empty
\bibitem [{\citenamefont {Henshaw}\ and\ \citenamefont
  {Woods}(1961)}]{Henshaw1961}%
  \BibitemOpen
  \bibfield  {author} {\bibinfo {author} {\bibfnamefont {D.~G.}\ \bibnamefont
  {Henshaw}}\ and\ \bibinfo {author} {\bibfnamefont {A.~D.~B.}\ \bibnamefont
  {Woods}},\ }\href {\doibase 10.1103/PhysRev.121.1266} {\bibfield  {journal}
  {\bibinfo  {journal} {Phys. Rev.}\ }\textbf {\bibinfo {volume} {121}},\
  \bibinfo {pages} {1266} (\bibinfo {year} {1961})}\BibitemShut {NoStop}%
\bibitem [{\citenamefont {Kohn}(1959)}]{Kohn1959}%
  \BibitemOpen
  \bibfield  {author} {\bibinfo {author} {\bibfnamefont {W.}~\bibnamefont
  {Kohn}},\ }\href {\doibase 10.1103/PhysRevLett.2.393} {\bibfield  {journal}
  {\bibinfo  {journal} {Phys. Rev. Lett.}\ }\textbf {\bibinfo {volume} {2}},\
  \bibinfo {pages} {393} (\bibinfo {year} {1959})}\BibitemShut {NoStop}%
\bibitem [{\citenamefont {Renker}\ \emph {et~al.}(1973)\citenamefont {Renker},
  \citenamefont {Rietschel}, \citenamefont {Pintschovius}, \citenamefont
  {Gl\"aser}, \citenamefont {Br\"uesch}, \citenamefont {Kuse},\ and\
  \citenamefont {Rice}}]{Renker1973}%
  \BibitemOpen
  \bibfield  {author} {\bibinfo {author} {\bibfnamefont {B.}~\bibnamefont
  {Renker}}, \bibinfo {author} {\bibfnamefont {H.}~\bibnamefont {Rietschel}},
  \bibinfo {author} {\bibfnamefont {L.}~\bibnamefont {Pintschovius}}, \bibinfo
  {author} {\bibfnamefont {W.}~\bibnamefont {Gl\"aser}}, \bibinfo {author}
  {\bibfnamefont {P.}~\bibnamefont {Br\"uesch}}, \bibinfo {author}
  {\bibfnamefont {D.}~\bibnamefont {Kuse}}, \ and\ \bibinfo {author}
  {\bibfnamefont {M.~J.}\ \bibnamefont {Rice}},\ }\href {\doibase
  10.1103/PhysRevLett.30.1144} {\bibfield  {journal} {\bibinfo  {journal}
  {Phys. Rev. Lett.}\ }\textbf {\bibinfo {volume} {30}},\ \bibinfo {pages}
  {1144} (\bibinfo {year} {1973})}\BibitemShut {NoStop}%
\bibitem [{\citenamefont {Godfrin}\ \emph {et~al.}(2012)\citenamefont
  {Godfrin}, \citenamefont {Meschke}, \citenamefont {Lauter}, \citenamefont
  {Sultan}, \citenamefont {B{\"o}hm}, \citenamefont {Krotscheck},\ and\
  \citenamefont {Panholzer}}]{Godfrin2012}%
  \BibitemOpen
  \bibfield  {author} {\bibinfo {author} {\bibfnamefont {H.}~\bibnamefont
  {Godfrin}}, \bibinfo {author} {\bibfnamefont {M.}~\bibnamefont {Meschke}},
  \bibinfo {author} {\bibfnamefont {H.-J.}\ \bibnamefont {Lauter}}, \bibinfo
  {author} {\bibfnamefont {A.}~\bibnamefont {Sultan}}, \bibinfo {author}
  {\bibfnamefont {H.~M.}\ \bibnamefont {B{\"o}hm}}, \bibinfo {author}
  {\bibfnamefont {E.}~\bibnamefont {Krotscheck}}, \ and\ \bibinfo {author}
  {\bibfnamefont {M.}~\bibnamefont {Panholzer}},\ }\href {\doibase
  10.1038/nature10919} {\bibfield  {journal} {\bibinfo  {journal} {Nature}\
  }\textbf {\bibinfo {volume} {483}},\ \bibinfo {pages} {576} (\bibinfo {year}
  {2012})}\BibitemShut {NoStop}%
\bibitem [{\citenamefont {Chomaz}\ \emph {et~al.}(2018)\citenamefont {Chomaz},
  \citenamefont {van Bijnen}, \citenamefont {Petter}, \citenamefont {Faraoni},
  \citenamefont {Baier}, \citenamefont {Becher}, \citenamefont {Mark},
  \citenamefont {W{\"a}chtler}, \citenamefont {Santos},\ and\ \citenamefont
  {Ferlaino}}]{Chomaz2018}%
  \BibitemOpen
  \bibfield  {author} {\bibinfo {author} {\bibfnamefont {L.}~\bibnamefont
  {Chomaz}}, \bibinfo {author} {\bibfnamefont {R.~M.~W.}\ \bibnamefont {van
  Bijnen}}, \bibinfo {author} {\bibfnamefont {D.}~\bibnamefont {Petter}},
  \bibinfo {author} {\bibfnamefont {G.}~\bibnamefont {Faraoni}}, \bibinfo
  {author} {\bibfnamefont {S.}~\bibnamefont {Baier}}, \bibinfo {author}
  {\bibfnamefont {J.~H.}\ \bibnamefont {Becher}}, \bibinfo {author}
  {\bibfnamefont {M.~J.}\ \bibnamefont {Mark}}, \bibinfo {author}
  {\bibfnamefont {F.}~\bibnamefont {W{\"a}chtler}}, \bibinfo {author}
  {\bibfnamefont {L.}~\bibnamefont {Santos}}, \ and\ \bibinfo {author}
  {\bibfnamefont {F.}~\bibnamefont {Ferlaino}},\ }\href {\doibase
  10.1038/s41567-018-0054-7} {\bibfield  {journal} {\bibinfo  {journal} {Nature
  Physics}\ }\textbf {\bibinfo {volume} {14}},\ \bibinfo {pages} {442}
  (\bibinfo {year} {2018})}\BibitemShut {NoStop}%
\bibitem [{\citenamefont {Kishine}\ \emph {et~al.}(2020)\citenamefont
  {Kishine}, \citenamefont {Ovchinnikov},\ and\ \citenamefont
  {Tereshchenko}}]{Kishine2020}%
  \BibitemOpen
  \bibfield  {author} {\bibinfo {author} {\bibfnamefont {J.}~\bibnamefont
  {Kishine}}, \bibinfo {author} {\bibfnamefont {A.~S.}\ \bibnamefont
  {Ovchinnikov}}, \ and\ \bibinfo {author} {\bibfnamefont {A.~A.}\ \bibnamefont
  {Tereshchenko}},\ }\href {\doibase 10.1103/PhysRevLett.125.245302} {\bibfield
   {journal} {\bibinfo  {journal} {Phys. Rev. Lett.}\ }\textbf {\bibinfo
  {volume} {125}},\ \bibinfo {pages} {245302} (\bibinfo {year}
  {2020})}\BibitemShut {NoStop}%
\bibitem [{\citenamefont {Chen}\ \emph {et~al.}(2021)\citenamefont {Chen},
  \citenamefont {Kadic},\ and\ \citenamefont {Wegener}}]{Chen2021}%
  \BibitemOpen
  \bibfield  {author} {\bibinfo {author} {\bibfnamefont {Y.}~\bibnamefont
  {Chen}}, \bibinfo {author} {\bibfnamefont {M.}~\bibnamefont {Kadic}}, \ and\
  \bibinfo {author} {\bibfnamefont {M.}~\bibnamefont {Wegener}},\ }\href
  {\doibase 10.1038/s41467-021-23574-2} {\bibfield  {journal} {\bibinfo
  {journal} {Nature Communications}\ }\textbf {\bibinfo {volume} {12}},\
  \bibinfo {pages} {3278} (\bibinfo {year} {2021})}\BibitemShut {NoStop}%
\bibitem [{\citenamefont {M\"uller}\ and\ \citenamefont
  {Burkard}(1979)}]{Muller1979}%
  \BibitemOpen
  \bibfield  {author} {\bibinfo {author} {\bibfnamefont {K.~A.}\ \bibnamefont
  {M\"uller}}\ and\ \bibinfo {author} {\bibfnamefont {H.}~\bibnamefont
  {Burkard}},\ }\href {\doibase 10.1103/PhysRevB.19.3593} {\bibfield  {journal}
  {\bibinfo  {journal} {Phys. Rev. B}\ }\textbf {\bibinfo {volume} {19}},\
  \bibinfo {pages} {3593} (\bibinfo {year} {1979})}\BibitemShut {NoStop}%
\bibitem [{\citenamefont {Schneider}\ \emph {et~al.}(1976)\citenamefont
  {Schneider}, \citenamefont {Beck},\ and\ \citenamefont
  {Stoll}}]{Schneider1976}%
  \BibitemOpen
  \bibfield  {author} {\bibinfo {author} {\bibfnamefont {T.}~\bibnamefont
  {Schneider}}, \bibinfo {author} {\bibfnamefont {H.}~\bibnamefont {Beck}}, \
  and\ \bibinfo {author} {\bibfnamefont {E.}~\bibnamefont {Stoll}},\ }\href
  {\doibase 10.1103/PhysRevB.13.1123} {\bibfield  {journal} {\bibinfo
  {journal} {Phys. Rev. B}\ }\textbf {\bibinfo {volume} {13}},\ \bibinfo
  {pages} {1123} (\bibinfo {year} {1976})}\BibitemShut {NoStop}%
\bibitem [{\citenamefont {Shin}\ \emph {et~al.}(2021)\citenamefont {Shin},
  \citenamefont {Latini}, \citenamefont {Sch\"afer}, \citenamefont {Sato},
  \citenamefont {De~Giovannini}, \citenamefont {H\"ubener},\ and\ \citenamefont
  {Rubio}}]{Shin2021}%
  \BibitemOpen
  \bibfield  {author} {\bibinfo {author} {\bibfnamefont {D.}~\bibnamefont
  {Shin}}, \bibinfo {author} {\bibfnamefont {S.}~\bibnamefont {Latini}},
  \bibinfo {author} {\bibfnamefont {C.}~\bibnamefont {Sch\"afer}}, \bibinfo
  {author} {\bibfnamefont {S.~A.}\ \bibnamefont {Sato}}, \bibinfo {author}
  {\bibfnamefont {U.}~\bibnamefont {De~Giovannini}}, \bibinfo {author}
  {\bibfnamefont {H.}~\bibnamefont {H\"ubener}}, \ and\ \bibinfo {author}
  {\bibfnamefont {A.}~\bibnamefont {Rubio}},\ }\href {\doibase
  10.1103/PhysRevB.104.L060103} {\bibfield  {journal} {\bibinfo  {journal}
  {Phys. Rev. B}\ }\textbf {\bibinfo {volume} {104}},\ \bibinfo {pages}
  {L060103} (\bibinfo {year} {2021})}\BibitemShut {NoStop}%
\bibitem [{\citenamefont {Esswein}\ and\ \citenamefont
  {Spaldin}(2021)}]{esswein2021}%
  \BibitemOpen
  \bibfield  {author} {\bibinfo {author} {\bibfnamefont {T.}~\bibnamefont
  {Esswein}}\ and\ \bibinfo {author} {\bibfnamefont {N.~A.}\ \bibnamefont
  {Spaldin}},\ }\href@noop {} {\enquote {\bibinfo {title} {Ferroelectric,
  quantum paraelectric or paraelectric? calculating the evolution from
  batio$_3$ to srtio$_3$ to ktao$_3$ using a single-particle quantum-mechanical
  description of the ions},}\ } (\bibinfo {year} {2021}),\ \Eprint
  {http://arxiv.org/abs/2112.11284} {arXiv:2112.11284 [cond-mat.mtrl-sci]}
  \BibitemShut {NoStop}%
\bibitem [{\citenamefont {Zubko}\ \emph {et~al.}(2011)\citenamefont {Zubko},
  \citenamefont {Gariglio}, \citenamefont {Gabay}, \citenamefont {Ghosez},\
  and\ \citenamefont {Triscone}}]{Zubko2011}%
  \BibitemOpen
  \bibfield  {author} {\bibinfo {author} {\bibfnamefont {P.}~\bibnamefont
  {Zubko}}, \bibinfo {author} {\bibfnamefont {S.}~\bibnamefont {Gariglio}},
  \bibinfo {author} {\bibfnamefont {M.}~\bibnamefont {Gabay}}, \bibinfo
  {author} {\bibfnamefont {P.}~\bibnamefont {Ghosez}}, \ and\ \bibinfo {author}
  {\bibfnamefont {J.-M.}\ \bibnamefont {Triscone}},\ }\href {\doibase
  10.1146/annurev-conmatphys-062910-140445} {\bibfield  {journal} {\bibinfo
  {journal} {Annual Review of Condensed Matter Physics}\ }\textbf {\bibinfo
  {volume} {2}},\ \bibinfo {pages} {141} (\bibinfo {year} {2011})},\ \Eprint
  {http://arxiv.org/abs/https://doi.org/10.1146/annurev-conmatphys-062910-140445}
  {https://doi.org/10.1146/annurev-conmatphys-062910-140445} \BibitemShut
  {NoStop}%
\bibitem [{\citenamefont {Collignon}\ \emph {et~al.}(2019)\citenamefont
  {Collignon}, \citenamefont {Lin}, \citenamefont {Rischau}, \citenamefont
  {Fauqu\'e},\ and\ \citenamefont {Behnia}}]{CollignonRev2019}%
  \BibitemOpen
  \bibfield  {author} {\bibinfo {author} {\bibfnamefont {C.}~\bibnamefont
  {Collignon}}, \bibinfo {author} {\bibfnamefont {X.}~\bibnamefont {Lin}},
  \bibinfo {author} {\bibfnamefont {C.~W.}\ \bibnamefont {Rischau}}, \bibinfo
  {author} {\bibfnamefont {B.}~\bibnamefont {Fauqu\'e}}, \ and\ \bibinfo
  {author} {\bibfnamefont {K.}~\bibnamefont {Behnia}},\ }\href {\doibase
  10.1146/annurev-conmatphys-031218-013144} {\bibfield  {journal} {\bibinfo
  {journal} {Annual Review of Condensed Matter Physics}\ }\textbf {\bibinfo
  {volume} {10}},\ \bibinfo {pages} {25} (\bibinfo {year} {2019})}\BibitemShut
  {NoStop}%
\bibitem [{\citenamefont {Rowley}\ \emph {et~al.}(2014)\citenamefont {Rowley},
  \citenamefont {Spalek}, \citenamefont {Smith}, \citenamefont {Dean},
  \citenamefont {Itoh}, \citenamefont {Scott}, \citenamefont {Lonzarich},\ and\
  \citenamefont {Saxena}}]{Rowley2014}%
  \BibitemOpen
  \bibfield  {author} {\bibinfo {author} {\bibfnamefont {S.~E.}\ \bibnamefont
  {Rowley}}, \bibinfo {author} {\bibfnamefont {L.~J.}\ \bibnamefont {Spalek}},
  \bibinfo {author} {\bibfnamefont {R.~P.}\ \bibnamefont {Smith}}, \bibinfo
  {author} {\bibfnamefont {M.~P.~M.}\ \bibnamefont {Dean}}, \bibinfo {author}
  {\bibfnamefont {M.}~\bibnamefont {Itoh}}, \bibinfo {author} {\bibfnamefont
  {J.~F.}\ \bibnamefont {Scott}}, \bibinfo {author} {\bibfnamefont {G.~G.}\
  \bibnamefont {Lonzarich}}, \ and\ \bibinfo {author} {\bibfnamefont {S.~S.}\
  \bibnamefont {Saxena}},\ }\href {\doibase 10.1038/nphys2924} {\bibfield
  {journal} {\bibinfo  {journal} {Nature Physics}\ }\textbf {\bibinfo {volume}
  {10}},\ \bibinfo {pages} {367} (\bibinfo {year} {2014})}\BibitemShut
  {NoStop}%
\bibitem [{\citenamefont {Coak}\ \emph {et~al.}(2020)\citenamefont {Coak},
  \citenamefont {Haines}, \citenamefont {Liu}, \citenamefont {Rowley},
  \citenamefont {Lonzarich},\ and\ \citenamefont {Saxena}}]{Coak12707}%
  \BibitemOpen
  \bibfield  {author} {\bibinfo {author} {\bibfnamefont {M.~J.}\ \bibnamefont
  {Coak}}, \bibinfo {author} {\bibfnamefont {C.~R.~S.}\ \bibnamefont {Haines}},
  \bibinfo {author} {\bibfnamefont {C.}~\bibnamefont {Liu}}, \bibinfo {author}
  {\bibfnamefont {S.~E.}\ \bibnamefont {Rowley}}, \bibinfo {author}
  {\bibfnamefont {G.~G.}\ \bibnamefont {Lonzarich}}, \ and\ \bibinfo {author}
  {\bibfnamefont {S.~S.}\ \bibnamefont {Saxena}},\ }\href {\doibase
  10.1073/pnas.1922151117} {\bibfield  {journal} {\bibinfo  {journal}
  {Proceedings of the National Academy of Sciences}\ }\textbf {\bibinfo
  {volume} {117}},\ \bibinfo {pages} {12707} (\bibinfo {year} {2020})},\
  \Eprint
  {http://arxiv.org/abs/https://www.pnas.org/content/117/23/12707.full.pdf}
  {https://www.pnas.org/content/117/23/12707.full.pdf} \BibitemShut {NoStop}%
\bibitem [{\citenamefont {Chandra}\ \emph {et~al.}(2017)\citenamefont
  {Chandra}, \citenamefont {Lonzarich}, \citenamefont {Rowley},\ and\
  \citenamefont {Scott}}]{Chandra_2017}%
  \BibitemOpen
  \bibfield  {author} {\bibinfo {author} {\bibfnamefont {P.}~\bibnamefont
  {Chandra}}, \bibinfo {author} {\bibfnamefont {G.~G.}\ \bibnamefont
  {Lonzarich}}, \bibinfo {author} {\bibfnamefont {S.~E.}\ \bibnamefont
  {Rowley}}, \ and\ \bibinfo {author} {\bibfnamefont {J.~F.}\ \bibnamefont
  {Scott}},\ }\href {\doibase 10.1088/1361-6633/aa82d2} {\bibfield  {journal}
  {\bibinfo  {journal} {Reports on Progress in Physics}\ }\textbf {\bibinfo
  {volume} {80}},\ \bibinfo {pages} {112502} (\bibinfo {year}
  {2017})}\BibitemShut {NoStop}%
\bibitem [{\citenamefont {Lin}\ \emph {et~al.}(2013)\citenamefont {Lin},
  \citenamefont {Zhu}, \citenamefont {Fauqu\'e},\ and\ \citenamefont
  {Behnia}}]{Lin2013}%
  \BibitemOpen
  \bibfield  {author} {\bibinfo {author} {\bibfnamefont {X.}~\bibnamefont
  {Lin}}, \bibinfo {author} {\bibfnamefont {Z.}~\bibnamefont {Zhu}}, \bibinfo
  {author} {\bibfnamefont {B.}~\bibnamefont {Fauqu\'e}}, \ and\ \bibinfo
  {author} {\bibfnamefont {K.}~\bibnamefont {Behnia}},\ }\href {\doibase
  10.1103/PhysRevX.3.021002} {\bibfield  {journal} {\bibinfo  {journal} {Phys.
  Rev. X}\ }\textbf {\bibinfo {volume} {3}},\ \bibinfo {pages} {021002}
  (\bibinfo {year} {2013})}\BibitemShut {NoStop}%
\bibitem [{\citenamefont {Rischau}\ \emph {et~al.}(2017)\citenamefont
  {Rischau}, \citenamefont {Lin}, \citenamefont {Grams}, \citenamefont {Finck},
  \citenamefont {Harms}, \citenamefont {Engelmayer}, \citenamefont {Lorenz},
  \citenamefont {Gallais}, \citenamefont {Fauqu{\'e}}, \citenamefont
  {Hemberger},\ and\ \citenamefont {Behnia}}]{Rischau2017}%
  \BibitemOpen
  \bibfield  {author} {\bibinfo {author} {\bibfnamefont {C.~W.}\ \bibnamefont
  {Rischau}}, \bibinfo {author} {\bibfnamefont {X.}~\bibnamefont {Lin}},
  \bibinfo {author} {\bibfnamefont {C.~P.}\ \bibnamefont {Grams}}, \bibinfo
  {author} {\bibfnamefont {D.}~\bibnamefont {Finck}}, \bibinfo {author}
  {\bibfnamefont {S.}~\bibnamefont {Harms}}, \bibinfo {author} {\bibfnamefont
  {J.}~\bibnamefont {Engelmayer}}, \bibinfo {author} {\bibfnamefont
  {T.}~\bibnamefont {Lorenz}}, \bibinfo {author} {\bibfnamefont
  {Y.}~\bibnamefont {Gallais}}, \bibinfo {author} {\bibfnamefont
  {B.}~\bibnamefont {Fauqu{\'e}}}, \bibinfo {author} {\bibfnamefont
  {J.}~\bibnamefont {Hemberger}}, \ and\ \bibinfo {author} {\bibfnamefont
  {K.}~\bibnamefont {Behnia}},\ }\href {https://doi.org/10.1038/nphys4085}
  {\bibfield  {journal} {\bibinfo  {journal} {Nature Physics}\ }\textbf
  {\bibinfo {volume} {13}},\ \bibinfo {pages} {643 EP } (\bibinfo {year}
  {2017})}\BibitemShut {NoStop}%
\bibitem [{\citenamefont {Ahadi}\ \emph {et~al.}(2019)\citenamefont {Ahadi},
  \citenamefont {Galletti}, \citenamefont {Li}, \citenamefont {Salmani-Rezaie},
  \citenamefont {Wu},\ and\ \citenamefont {Stemmer}}]{Ahadi2019}%
  \BibitemOpen
  \bibfield  {author} {\bibinfo {author} {\bibfnamefont {K.}~\bibnamefont
  {Ahadi}}, \bibinfo {author} {\bibfnamefont {L.}~\bibnamefont {Galletti}},
  \bibinfo {author} {\bibfnamefont {Y.}~\bibnamefont {Li}}, \bibinfo {author}
  {\bibfnamefont {S.}~\bibnamefont {Salmani-Rezaie}}, \bibinfo {author}
  {\bibfnamefont {W.}~\bibnamefont {Wu}}, \ and\ \bibinfo {author}
  {\bibfnamefont {S.}~\bibnamefont {Stemmer}},\ }\href {\doibase
  10.1126/sciadv.aaw0120} {\bibfield  {journal} {\bibinfo  {journal} {Science
  Advances}\ }\textbf {\bibinfo {volume} {5}},\ \bibinfo {pages} {eaaw0120}
  (\bibinfo {year} {2019})},\ \Eprint
  {http://arxiv.org/abs/https://www.science.org/doi/pdf/10.1126/sciadv.aaw0120}
  {https://www.science.org/doi/pdf/10.1126/sciadv.aaw0120} \BibitemShut
  {NoStop}%
\bibitem [{\citenamefont {Lin}\ \emph {et~al.}(2015)\citenamefont {Lin},
  \citenamefont {Fauqu{\'e}},\ and\ \citenamefont {Behnia}}]{Lin2015sc}%
  \BibitemOpen
  \bibfield  {author} {\bibinfo {author} {\bibfnamefont {X.}~\bibnamefont
  {Lin}}, \bibinfo {author} {\bibfnamefont {B.}~\bibnamefont {Fauqu{\'e}}}, \
  and\ \bibinfo {author} {\bibfnamefont {K.}~\bibnamefont {Behnia}},\ }\href
  {\doibase 10.1126/science.aaa8655} {\bibfield  {journal} {\bibinfo  {journal}
  {Science}\ }\textbf {\bibinfo {volume} {349}},\ \bibinfo {pages} {945}
  (\bibinfo {year} {2015})}\BibitemShut {NoStop}%
\bibitem [{\citenamefont {Edge}\ \emph {et~al.}(2015)\citenamefont {Edge},
  \citenamefont {Kedem}, \citenamefont {Aschauer}, \citenamefont {Spaldin},\
  and\ \citenamefont {Balatsky}}]{Edge2015}%
  \BibitemOpen
  \bibfield  {author} {\bibinfo {author} {\bibfnamefont {J.~M.}\ \bibnamefont
  {Edge}}, \bibinfo {author} {\bibfnamefont {Y.}~\bibnamefont {Kedem}},
  \bibinfo {author} {\bibfnamefont {U.}~\bibnamefont {Aschauer}}, \bibinfo
  {author} {\bibfnamefont {N.~A.}\ \bibnamefont {Spaldin}}, \ and\ \bibinfo
  {author} {\bibfnamefont {A.~V.}\ \bibnamefont {Balatsky}},\ }\href {\doibase
  10.1103/PhysRevLett.115.247002} {\bibfield  {journal} {\bibinfo  {journal}
  {Phys. Rev. Lett.}\ }\textbf {\bibinfo {volume} {115}},\ \bibinfo {pages}
  {247002} (\bibinfo {year} {2015})}\BibitemShut {NoStop}%
\bibitem [{\citenamefont {van~der Marel}\ \emph {et~al.}(2019)\citenamefont
  {van~der Marel}, \citenamefont {Barantani},\ and\ \citenamefont
  {Rischau}}]{Marel2019}%
  \BibitemOpen
  \bibfield  {author} {\bibinfo {author} {\bibfnamefont {D.}~\bibnamefont
  {van~der Marel}}, \bibinfo {author} {\bibfnamefont {F.}~\bibnamefont
  {Barantani}}, \ and\ \bibinfo {author} {\bibfnamefont {C.~W.}\ \bibnamefont
  {Rischau}},\ }\href {\doibase 10.1103/PhysRevResearch.1.013003} {\bibfield
  {journal} {\bibinfo  {journal} {Phys. Rev. Research}\ }\textbf {\bibinfo
  {volume} {1}},\ \bibinfo {pages} {013003} (\bibinfo {year}
  {2019})}\BibitemShut {NoStop}%
\bibitem [{\citenamefont {Collignon}\ \emph {et~al.}(2020)\citenamefont
  {Collignon}, \citenamefont {Bourges}, \citenamefont {Fauqu\'e},\ and\
  \citenamefont {Behnia}}]{Collignon2020}%
  \BibitemOpen
  \bibfield  {author} {\bibinfo {author} {\bibfnamefont {C.}~\bibnamefont
  {Collignon}}, \bibinfo {author} {\bibfnamefont {P.}~\bibnamefont {Bourges}},
  \bibinfo {author} {\bibfnamefont {B.}~\bibnamefont {Fauqu\'e}}, \ and\
  \bibinfo {author} {\bibfnamefont {K.}~\bibnamefont {Behnia}},\ }\href
  {\doibase 10.1103/PhysRevX.10.031025} {\bibfield  {journal} {\bibinfo
  {journal} {Phys. Rev. X}\ }\textbf {\bibinfo {volume} {10}},\ \bibinfo
  {pages} {031025} (\bibinfo {year} {2020})}\BibitemShut {NoStop}%
\bibitem [{\citenamefont {Collignon}\ \emph {et~al.}(2021)\citenamefont
  {Collignon}, \citenamefont {Awashima}, \citenamefont {Ravi}, \citenamefont
  {Lin}, \citenamefont {Rischau}, \citenamefont {Acheche}, \citenamefont
  {Vignolle}, \citenamefont {Proust}, \citenamefont {Fuseya}, \citenamefont
  {Behnia},\ and\ \citenamefont {Fauqu\'e}}]{Collignon2021}%
  \BibitemOpen
  \bibfield  {author} {\bibinfo {author} {\bibfnamefont {C.}~\bibnamefont
  {Collignon}}, \bibinfo {author} {\bibfnamefont {Y.}~\bibnamefont {Awashima}},
  \bibinfo {author} {\bibnamefont {Ravi}}, \bibinfo {author} {\bibfnamefont
  {X.}~\bibnamefont {Lin}}, \bibinfo {author} {\bibfnamefont {C.~W.}\
  \bibnamefont {Rischau}}, \bibinfo {author} {\bibfnamefont {A.}~\bibnamefont
  {Acheche}}, \bibinfo {author} {\bibfnamefont {B.}~\bibnamefont {Vignolle}},
  \bibinfo {author} {\bibfnamefont {C.}~\bibnamefont {Proust}}, \bibinfo
  {author} {\bibfnamefont {Y.}~\bibnamefont {Fuseya}}, \bibinfo {author}
  {\bibfnamefont {K.}~\bibnamefont {Behnia}}, \ and\ \bibinfo {author}
  {\bibfnamefont {B.}~\bibnamefont {Fauqu\'e}},\ }\href {\doibase
  10.1103/PhysRevMaterials.5.065002} {\bibfield  {journal} {\bibinfo  {journal}
  {Phys. Rev. Materials}\ }\textbf {\bibinfo {volume} {5}},\ \bibinfo {pages}
  {065002} (\bibinfo {year} {2021})}\BibitemShut {NoStop}%
\bibitem [{\citenamefont {Kumar}\ \emph {et~al.}(2021)\citenamefont {Kumar},
  \citenamefont {Yudson},\ and\ \citenamefont {Maslov}}]{Kumar2021}%
  \BibitemOpen
  \bibfield  {author} {\bibinfo {author} {\bibfnamefont {A.}~\bibnamefont
  {Kumar}}, \bibinfo {author} {\bibfnamefont {V.~I.}\ \bibnamefont {Yudson}}, \
  and\ \bibinfo {author} {\bibfnamefont {D.~L.}\ \bibnamefont {Maslov}},\
  }\href {\doibase 10.1103/PhysRevLett.126.076601} {\bibfield  {journal}
  {\bibinfo  {journal} {Phys. Rev. Lett.}\ }\textbf {\bibinfo {volume} {126}},\
  \bibinfo {pages} {076601} (\bibinfo {year} {2021})}\BibitemShut {NoStop}%
\bibitem [{\citenamefont {Kiseliov}\ and\ \citenamefont
  {Feigel'man}(2021)}]{kiseliov2021theory}%
  \BibitemOpen
  \bibfield  {author} {\bibinfo {author} {\bibfnamefont {D.}~\bibnamefont
  {Kiseliov}}\ and\ \bibinfo {author} {\bibfnamefont {M.}~\bibnamefont
  {Feigel'man}},\ }\href@noop {} {\enquote {\bibinfo {title} {Theory of
  superconductivity due to ngai's mechanism in lightly doped srtio3},}\ }
  (\bibinfo {year} {2021}),\ \Eprint {http://arxiv.org/abs/2106.09530}
  {arXiv:2106.09530 [cond-mat.supr-con]} \BibitemShut {NoStop}%
\bibitem [{\citenamefont {M{\"u}ller}\ \emph {et~al.}(1991)\citenamefont
  {M{\"u}ller}, \citenamefont {Berlinger},\ and\ \citenamefont
  {Tosatti}}]{Muller1991}%
  \BibitemOpen
  \bibfield  {author} {\bibinfo {author} {\bibfnamefont {K.~A.}\ \bibnamefont
  {M{\"u}ller}}, \bibinfo {author} {\bibfnamefont {W.}~\bibnamefont
  {Berlinger}}, \ and\ \bibinfo {author} {\bibfnamefont {E.}~\bibnamefont
  {Tosatti}},\ }\href {\doibase 10.1007/BF01313549} {\bibfield  {journal}
  {\bibinfo  {journal} {Zeitschrift f{\"u}r Physik B Condensed Matter}\
  }\textbf {\bibinfo {volume} {84}},\ \bibinfo {pages} {277} (\bibinfo {year}
  {1991})}\BibitemShut {NoStop}%
\bibitem [{\citenamefont {Lemanov}(2002)}]{Lemanov2002}%
  \BibitemOpen
  \bibfield  {author} {\bibinfo {author} {\bibfnamefont {V.~V.}\ \bibnamefont
  {Lemanov}},\ }\href {\doibase 10.1080/00150190208260600} {\bibfield
  {journal} {\bibinfo  {journal} {Ferroelectrics}\ }\textbf {\bibinfo {volume}
  {265}},\ \bibinfo {pages} {1} (\bibinfo {year} {2002})},\ \Eprint
  {http://arxiv.org/abs/https://doi.org/10.1080/00150190208260600}
  {https://doi.org/10.1080/00150190208260600} \BibitemShut {NoStop}%
\bibitem [{SM()}]{SM}%
  \BibitemOpen
  \href@noop {} {}\bibinfo {howpublished} {See Supplemental Material for the
  presentation of the samples, the spectrometer configurations and the
  experimentals fitting procedure and the raw datas of Fig. 1
  \cite{WEBER,Yamanaka2000}}\BibitemShut {NoStop}%
\bibitem [{\citenamefont {Shirane}\ and\ \citenamefont
  {Yamada}(1969)}]{Shirane1969}%
  \BibitemOpen
  \bibfield  {author} {\bibinfo {author} {\bibfnamefont {G.}~\bibnamefont
  {Shirane}}\ and\ \bibinfo {author} {\bibfnamefont {Y.}~\bibnamefont
  {Yamada}},\ }\href {\doibase 10.1103/PhysRev.177.858} {\bibfield  {journal}
  {\bibinfo  {journal} {Phys. Rev.}\ }\textbf {\bibinfo {volume} {177}},\
  \bibinfo {pages} {858} (\bibinfo {year} {1969})}\BibitemShut {NoStop}%
\bibitem [{\citenamefont {Yamada}\ and\ \citenamefont
  {Shirane}(1969)}]{Yamada1969}%
  \BibitemOpen
  \bibfield  {author} {\bibinfo {author} {\bibfnamefont {Y.}~\bibnamefont
  {Yamada}}\ and\ \bibinfo {author} {\bibfnamefont {G.}~\bibnamefont
  {Shirane}},\ }\href {\doibase 10.1143/JPSJ.26.396} {\bibfield  {journal}
  {\bibinfo  {journal} {Journal of the Physical Society of Japan}\ }\textbf
  {\bibinfo {volume} {26}},\ \bibinfo {pages} {396} (\bibinfo {year} {1969})},\
  \Eprint {http://arxiv.org/abs/https://doi.org/10.1143/JPSJ.26.396}
  {https://doi.org/10.1143/JPSJ.26.396} \BibitemShut {NoStop}%
\bibitem [{\citenamefont {Fleury}\ \emph {et~al.}(1968)\citenamefont {Fleury},
  \citenamefont {Scott},\ and\ \citenamefont {Worlock}}]{Fleury1968}%
  \BibitemOpen
  \bibfield  {author} {\bibinfo {author} {\bibfnamefont {P.~A.}\ \bibnamefont
  {Fleury}}, \bibinfo {author} {\bibfnamefont {J.~F.}\ \bibnamefont {Scott}}, \
  and\ \bibinfo {author} {\bibfnamefont {J.~M.}\ \bibnamefont {Worlock}},\
  }\href {\doibase 10.1103/PhysRevLett.21.16} {\bibfield  {journal} {\bibinfo
  {journal} {Phys. Rev. Lett.}\ }\textbf {\bibinfo {volume} {21}},\ \bibinfo
  {pages} {16} (\bibinfo {year} {1968})}\BibitemShut {NoStop}%
\bibitem [{\citenamefont {Vogt}(1995)}]{Vogt95}%
  \BibitemOpen
  \bibfield  {author} {\bibinfo {author} {\bibfnamefont {H.}~\bibnamefont
  {Vogt}},\ }\href {\doibase 10.1103/PhysRevB.51.8046} {\bibfield  {journal}
  {\bibinfo  {journal} {Phys. Rev. B}\ }\textbf {\bibinfo {volume} {51}},\
  \bibinfo {pages} {8046} (\bibinfo {year} {1995})}\BibitemShut {NoStop}%
\bibitem [{\citenamefont {Sirenko}\ \emph {et~al.}(2000)\citenamefont
  {Sirenko}, \citenamefont {Bernhard}, \citenamefont {Golnik}, \citenamefont
  {Clark}, \citenamefont {Hao}, \citenamefont {Si},\ and\ \citenamefont
  {Xi}}]{Sirenko2000}%
  \BibitemOpen
  \bibfield  {author} {\bibinfo {author} {\bibfnamefont {A.~A.}\ \bibnamefont
  {Sirenko}}, \bibinfo {author} {\bibfnamefont {C.}~\bibnamefont {Bernhard}},
  \bibinfo {author} {\bibfnamefont {A.}~\bibnamefont {Golnik}}, \bibinfo
  {author} {\bibfnamefont {A.~M.}\ \bibnamefont {Clark}}, \bibinfo {author}
  {\bibfnamefont {J.}~\bibnamefont {Hao}}, \bibinfo {author} {\bibfnamefont
  {W.}~\bibnamefont {Si}}, \ and\ \bibinfo {author} {\bibfnamefont {X.~X.}\
  \bibnamefont {Xi}},\ }\href {\doibase 10.1038/35006023} {\bibfield  {journal}
  {\bibinfo  {journal} {Nature}\ }\textbf {\bibinfo {volume} {404}},\ \bibinfo
  {pages} {373} (\bibinfo {year} {2000})}\BibitemShut {NoStop}%
\bibitem [{\citenamefont {van Mechelen}(2010)}]{vanMechelen2010}%
  \BibitemOpen
  \bibfield  {author} {\bibinfo {author} {\bibfnamefont {D.}~\bibnamefont {van
  Mechelen}},\ }\emph {\bibinfo {title} {{Charge and Spin Electrodynamics of
  SrTiO$_3$ and EuTiO$_3$ Studied by Optical Spectroscopy}}},\ \href
  {http://dqmp.unige.ch/vandermarel/wp-content/uploads/2016/05/thesis_vanmechelen.pdf}
  {Ph.D. thesis},\ \bibinfo  {school} {Geneva University} (\bibinfo {year}
  {2010})\BibitemShut {NoStop}%
\bibitem [{\citenamefont {Cowley}\ and\ \citenamefont
  {Salje}(1996)}]{Cowley1996}%
  \BibitemOpen
  \bibfield  {author} {\bibinfo {author} {\bibfnamefont {R.~A.}\ \bibnamefont
  {Cowley}}\ and\ \bibinfo {author} {\bibfnamefont {E.~K.~H.}\ \bibnamefont
  {Salje}},\ }\href {\doibase 10.1098/rsta.1996.0130} {\bibfield  {journal}
  {\bibinfo  {journal} {Philosophical Transactions of the Royal Society of
  London. Series A: Mathematical, Physical and Engineering Sciences}\ }\textbf
  {\bibinfo {volume} {354}},\ \bibinfo {pages} {2799} (\bibinfo {year}
  {1996})}\BibitemShut {NoStop}%
\bibitem [{\citenamefont {Balashova}\ \emph {et~al.}(1996)\citenamefont
  {Balashova}, \citenamefont {Lemanov}, \citenamefont {Kunze}, \citenamefont
  {Martin},\ and\ \citenamefont {Weihnacht}}]{Balashova1996}%
  \BibitemOpen
  \bibfield  {author} {\bibinfo {author} {\bibfnamefont {E.~V.}\ \bibnamefont
  {Balashova}}, \bibinfo {author} {\bibfnamefont {V.~V.}\ \bibnamefont
  {Lemanov}}, \bibinfo {author} {\bibfnamefont {R.}~\bibnamefont {Kunze}},
  \bibinfo {author} {\bibfnamefont {G.}~\bibnamefont {Martin}}, \ and\ \bibinfo
  {author} {\bibfnamefont {M.}~\bibnamefont {Weihnacht}},\ }\href {\doibase
  10.1080/00150199608224093} {\bibfield  {journal} {\bibinfo  {journal}
  {Ferroelectrics}\ }\textbf {\bibinfo {volume} {183}},\ \bibinfo {pages} {75}
  (\bibinfo {year} {1996})},\ \Eprint
  {http://arxiv.org/abs/https://doi.org/10.1080/00150199608224093}
  {https://doi.org/10.1080/00150199608224093} \BibitemShut {NoStop}%
\bibitem [{\citenamefont {Rehwald}(1970{\natexlab{a}})}]{Rehwald1970}%
  \BibitemOpen
  \bibfield  {author} {\bibinfo {author} {\bibfnamefont {W.}~\bibnamefont
  {Rehwald}},\ }\href {\doibase https://doi.org/10.1016/0038-1098(70)90725-8}
  {\bibfield  {journal} {\bibinfo  {journal} {Solid State Communications}\
  }\textbf {\bibinfo {volume} {8}},\ \bibinfo {pages} {1483} (\bibinfo {year}
  {1970}{\natexlab{a}})}\BibitemShut {NoStop}%
\bibitem [{\citenamefont {Rehwald}(1970{\natexlab{b}})}]{Rehwalda1970}%
  \BibitemOpen
  \bibfield  {author} {\bibinfo {author} {\bibfnamefont {W.}~\bibnamefont
  {Rehwald}},\ }\href {\doibase https://doi.org/10.1016/0038-1098(70)90159-6}
  {\bibfield  {journal} {\bibinfo  {journal} {Solid State Communications}\
  }\textbf {\bibinfo {volume} {8}},\ \bibinfo {pages} {607} (\bibinfo {year}
  {1970}{\natexlab{b}})}\BibitemShut {NoStop}%
\bibitem [{\citenamefont {Scott}\ and\ \citenamefont
  {Ledbetter}(1997)}]{Scott1997}%
  \BibitemOpen
  \bibfield  {author} {\bibinfo {author} {\bibfnamefont {J.}~\bibnamefont
  {Scott}}\ and\ \bibinfo {author} {\bibfnamefont {H.}~\bibnamefont
  {Ledbetter}},\ }\href {\doibase 10.1007/s002570050500} {\bibfield  {journal}
  {\bibinfo  {journal} {Zeitschrift für Physik B Condensed Matter}\ }\textbf
  {\bibinfo {volume} {104}},\ \bibinfo {pages} {635} (\bibinfo {year}
  {1997})}\BibitemShut {NoStop}%
\bibitem [{\citenamefont {L\"uthi}\ and\ \citenamefont
  {Moran}(1970)}]{Luthi1970}%
  \BibitemOpen
  \bibfield  {author} {\bibinfo {author} {\bibfnamefont {B.}~\bibnamefont
  {L\"uthi}}\ and\ \bibinfo {author} {\bibfnamefont {T.~J.}\ \bibnamefont
  {Moran}},\ }\href {\doibase 10.1103/PhysRevB.2.1211} {\bibfield  {journal}
  {\bibinfo  {journal} {Phys. Rev. B}\ }\textbf {\bibinfo {volume} {2}},\
  \bibinfo {pages} {1211} (\bibinfo {year} {1970})}\BibitemShut {NoStop}%
\bibitem [{\citenamefont {Laubereau}\ and\ \citenamefont
  {Zurek}(1970)}]{Laubereau1970}%
  \BibitemOpen
  \bibfield  {author} {\bibinfo {author} {\bibfnamefont {A.}~\bibnamefont
  {Laubereau}}\ and\ \bibinfo {author} {\bibfnamefont {R.}~\bibnamefont
  {Zurek}},\ }\href {\doibase doi:10.1515/zna-1970-0310} {\bibfield  {journal}
  {\bibinfo  {journal} {Zeitschrift f√ºr Naturforschung A}\ }\textbf
  {\bibinfo {volume} {25}},\ \bibinfo {pages} {391} (\bibinfo {year}
  {1970})}\BibitemShut {NoStop}%
\bibitem [{\citenamefont {Hehlen}\ \emph {et~al.}(1996)\citenamefont {Hehlen},
  \citenamefont {Kallassy},\ and\ \citenamefont {Courtens}}]{Hehlen1996}%
  \BibitemOpen
  \bibfield  {author} {\bibinfo {author} {\bibfnamefont {B.}~\bibnamefont
  {Hehlen}}, \bibinfo {author} {\bibfnamefont {Z.}~\bibnamefont {Kallassy}}, \
  and\ \bibinfo {author} {\bibfnamefont {E.}~\bibnamefont {Courtens}},\ }\href
  {\doibase 10.1080/00150199608224113} {\bibfield  {journal} {\bibinfo
  {journal} {Ferroelectrics}\ }\textbf {\bibinfo {volume} {183}},\ \bibinfo
  {pages} {265} (\bibinfo {year} {1996})},\ \Eprint
  {http://arxiv.org/abs/https://doi.org/10.1080/00150199608224113}
  {https://doi.org/10.1080/00150199608224113} \BibitemShut {NoStop}%
\bibitem [{\citenamefont {Lasota}\ \emph {et~al.}(1997)\citenamefont {Lasota},
  \citenamefont {Wang}, \citenamefont {Yu},\ and\ \citenamefont
  {Krakauer}}]{Lasota1997}%
  \BibitemOpen
  \bibfield  {author} {\bibinfo {author} {\bibfnamefont {C.}~\bibnamefont
  {Lasota}}, \bibinfo {author} {\bibfnamefont {C.-Z.}\ \bibnamefont {Wang}},
  \bibinfo {author} {\bibfnamefont {R.}~\bibnamefont {Yu}}, \ and\ \bibinfo
  {author} {\bibfnamefont {H.}~\bibnamefont {Krakauer}},\ }\href {\doibase
  10.1080/00150199708016086} {\bibfield  {journal} {\bibinfo  {journal}
  {Ferroelectrics}\ }\textbf {\bibinfo {volume} {194}},\ \bibinfo {pages} {109}
  (\bibinfo {year} {1997})}\BibitemShut {NoStop}%
\bibitem [{\citenamefont {Tadano}\ and\ \citenamefont
  {Tsuneyuki}(2015)}]{Tadano2015}%
  \BibitemOpen
  \bibfield  {author} {\bibinfo {author} {\bibfnamefont {T.}~\bibnamefont
  {Tadano}}\ and\ \bibinfo {author} {\bibfnamefont {S.}~\bibnamefont
  {Tsuneyuki}},\ }\href {\doibase 10.1103/PhysRevB.92.054301} {\bibfield
  {journal} {\bibinfo  {journal} {Phys. Rev. B}\ }\textbf {\bibinfo {volume}
  {92}},\ \bibinfo {pages} {054301} (\bibinfo {year} {2015})}\BibitemShut
  {NoStop}%
\bibitem [{\citenamefont {Zhou}\ \emph {et~al.}(2018)\citenamefont {Zhou},
  \citenamefont {Hellman},\ and\ \citenamefont {Bernardi}}]{Zhou2018}%
  \BibitemOpen
  \bibfield  {author} {\bibinfo {author} {\bibfnamefont {J.-J.}\ \bibnamefont
  {Zhou}}, \bibinfo {author} {\bibfnamefont {O.}~\bibnamefont {Hellman}}, \
  and\ \bibinfo {author} {\bibfnamefont {M.}~\bibnamefont {Bernardi}},\ }\href
  {\doibase 10.1103/PhysRevLett.121.226603} {\bibfield  {journal} {\bibinfo
  {journal} {Phys. Rev. Lett.}\ }\textbf {\bibinfo {volume} {121}},\ \bibinfo
  {pages} {226603} (\bibinfo {year} {2018})}\BibitemShut {NoStop}%
\bibitem [{\citenamefont {Fumega}\ \emph {et~al.}(2020)\citenamefont {Fumega},
  \citenamefont {Fu}, \citenamefont {Pardo},\ and\ \citenamefont
  {Singh}}]{Fumega2020}%
  \BibitemOpen
  \bibfield  {author} {\bibinfo {author} {\bibfnamefont {A.~O.}\ \bibnamefont
  {Fumega}}, \bibinfo {author} {\bibfnamefont {Y.}~\bibnamefont {Fu}}, \bibinfo
  {author} {\bibfnamefont {V.}~\bibnamefont {Pardo}}, \ and\ \bibinfo {author}
  {\bibfnamefont {D.~J.}\ \bibnamefont {Singh}},\ }\href {\doibase
  10.1103/PhysRevMaterials.4.033606} {\bibfield  {journal} {\bibinfo  {journal}
  {Phys. Rev. Materials}\ }\textbf {\bibinfo {volume} {4}},\ \bibinfo {pages}
  {033606} (\bibinfo {year} {2020})}\BibitemShut {NoStop}%
\bibitem [{\citenamefont {He}\ \emph {et~al.}(2020)\citenamefont {He},
  \citenamefont {Bansal}, \citenamefont {Winn}, \citenamefont {Chi},
  \citenamefont {Boatner},\ and\ \citenamefont {Delaire}}]{Delaire2020}%
  \BibitemOpen
  \bibfield  {author} {\bibinfo {author} {\bibfnamefont {X.}~\bibnamefont
  {He}}, \bibinfo {author} {\bibfnamefont {D.}~\bibnamefont {Bansal}}, \bibinfo
  {author} {\bibfnamefont {B.}~\bibnamefont {Winn}}, \bibinfo {author}
  {\bibfnamefont {S.}~\bibnamefont {Chi}}, \bibinfo {author} {\bibfnamefont
  {L.}~\bibnamefont {Boatner}}, \ and\ \bibinfo {author} {\bibfnamefont
  {O.}~\bibnamefont {Delaire}},\ }\href {\doibase
  10.1103/PhysRevLett.124.145901} {\bibfield  {journal} {\bibinfo  {journal}
  {Phys. Rev. Lett.}\ }\textbf {\bibinfo {volume} {124}},\ \bibinfo {pages}
  {145901} (\bibinfo {year} {2020})}\BibitemShut {NoStop}%
\bibitem [{\citenamefont {{van Roekeghem}}\ \emph {et~al.}(2021)\citenamefont
  {{van Roekeghem}}, \citenamefont {Carrete},\ and\ \citenamefont
  {Mingo}}]{vanRoekeghem2021}%
  \BibitemOpen
  \bibfield  {author} {\bibinfo {author} {\bibfnamefont {A.}~\bibnamefont {{van
  Roekeghem}}}, \bibinfo {author} {\bibfnamefont {J.}~\bibnamefont {Carrete}},
  \ and\ \bibinfo {author} {\bibfnamefont {N.}~\bibnamefont {Mingo}},\ }\href
  {\doibase https://doi.org/10.1016/j.cpc.2021.107945} {\bibfield  {journal}
  {\bibinfo  {journal} {Computer Physics Communications}\ }\textbf {\bibinfo
  {volume} {263}},\ \bibinfo {pages} {107945} (\bibinfo {year}
  {2021})}\BibitemShut {NoStop}%
\bibitem [{\citenamefont {Axe}\ \emph {et~al.}(1970)\citenamefont {Axe},
  \citenamefont {Harada},\ and\ \citenamefont {Shirane}}]{Axe1970}%
  \BibitemOpen
  \bibfield  {author} {\bibinfo {author} {\bibfnamefont {J.~D.}\ \bibnamefont
  {Axe}}, \bibinfo {author} {\bibfnamefont {J.}~\bibnamefont {Harada}}, \ and\
  \bibinfo {author} {\bibfnamefont {G.}~\bibnamefont {Shirane}},\ }\href
  {\doibase 10.1103/PhysRevB.1.1227} {\bibfield  {journal} {\bibinfo  {journal}
  {Phys. Rev. B}\ }\textbf {\bibinfo {volume} {1}},\ \bibinfo {pages} {1227}
  (\bibinfo {year} {1970})}\BibitemShut {NoStop}%
\bibitem [{\citenamefont {Hehlen}\ \emph {et~al.}(1998)\citenamefont {Hehlen},
  \citenamefont {Arzel}, \citenamefont {Tagantsev}, \citenamefont {Courtens},
  \citenamefont {Inaba}, \citenamefont {Yamanaka},\ and\ \citenamefont
  {Inoue}}]{Hehlen1998}%
  \BibitemOpen
  \bibfield  {author} {\bibinfo {author} {\bibfnamefont {B.}~\bibnamefont
  {Hehlen}}, \bibinfo {author} {\bibfnamefont {L.}~\bibnamefont {Arzel}},
  \bibinfo {author} {\bibfnamefont {A.~K.}\ \bibnamefont {Tagantsev}}, \bibinfo
  {author} {\bibfnamefont {E.}~\bibnamefont {Courtens}}, \bibinfo {author}
  {\bibfnamefont {Y.}~\bibnamefont {Inaba}}, \bibinfo {author} {\bibfnamefont
  {A.}~\bibnamefont {Yamanaka}}, \ and\ \bibinfo {author} {\bibfnamefont
  {K.}~\bibnamefont {Inoue}},\ }\href {\doibase 10.1103/PhysRevB.57.R13989}
  {\bibfield  {journal} {\bibinfo  {journal} {Phys. Rev. B}\ }\textbf {\bibinfo
  {volume} {57}},\ \bibinfo {pages} {R13989} (\bibinfo {year}
  {1998})}\BibitemShut {NoStop}%
\bibitem [{\citenamefont {Courtens}\ \emph {et~al.}(1993)\citenamefont
  {Courtens}, \citenamefont {Coddens}, \citenamefont {Hennion}, \citenamefont
  {Hehlen}, \citenamefont {Pelous},\ and\ \citenamefont
  {Vacher}}]{Courtens1993}%
  \BibitemOpen
  \bibfield  {author} {\bibinfo {author} {\bibfnamefont {E.}~\bibnamefont
  {Courtens}}, \bibinfo {author} {\bibfnamefont {G.}~\bibnamefont {Coddens}},
  \bibinfo {author} {\bibfnamefont {B.}~\bibnamefont {Hennion}}, \bibinfo
  {author} {\bibfnamefont {B.}~\bibnamefont {Hehlen}}, \bibinfo {author}
  {\bibfnamefont {J.}~\bibnamefont {Pelous}}, \ and\ \bibinfo {author}
  {\bibfnamefont {R.}~\bibnamefont {Vacher}},\ }\href {\doibase
  10.1088/0031-8949/1993/t49b/008} {\bibfield  {journal} {\bibinfo  {journal}
  {Physica Scripta}\ }\textbf {\bibinfo {volume} {T49B}},\ \bibinfo {pages}
  {430} (\bibinfo {year} {1993})}\BibitemShut {NoStop}%
\bibitem [{\citenamefont {Carpenter}(2007)}]{Carpenter2007}%
  \BibitemOpen
  \bibfield  {author} {\bibinfo {author} {\bibfnamefont {M.~A.}\ \bibnamefont
  {Carpenter}},\ }\href {\doibase 10.2138/am.2007.2295} {\bibfield  {journal}
  {\bibinfo  {journal} {American Mineralogist}\ }\textbf {\bibinfo {volume}
  {92}},\ \bibinfo {pages} {309} (\bibinfo {year} {2007})},\ \Eprint
  {http://arxiv.org/abs/https://pubs.geoscienceworld.org/msa/ammin/article-pdf/92/2-3/309/3620965/9\_2295Carpenter1.indd.pdf}
  {https://pubs.geoscienceworld.org/msa/ammin/article-pdf/92/2-3/309/3620965/9\_2295Carpenter1.indd.pdf}
  \BibitemShut {NoStop}%
\bibitem [{\citenamefont {Bussmann-Holder}\ \emph {et~al.}(2007)\citenamefont
  {Bussmann-Holder}, \citenamefont {B\"uttner},\ and\ \citenamefont
  {Bishop}}]{Bussmann-Holder2007}%
  \BibitemOpen
  \bibfield  {author} {\bibinfo {author} {\bibfnamefont {A.}~\bibnamefont
  {Bussmann-Holder}}, \bibinfo {author} {\bibfnamefont {H.}~\bibnamefont
  {B\"uttner}}, \ and\ \bibinfo {author} {\bibfnamefont {A.~R.}\ \bibnamefont
  {Bishop}},\ }\href {\doibase 10.1103/PhysRevLett.99.167603} {\bibfield
  {journal} {\bibinfo  {journal} {Phys. Rev. Lett.}\ }\textbf {\bibinfo
  {volume} {99}},\ \bibinfo {pages} {167603} (\bibinfo {year}
  {2007})}\BibitemShut {NoStop}%
\bibitem [{\citenamefont {Ktitorov}\ and\ \citenamefont
  {Jastrabik}(1998)}]{Ktitorov1998}%
  \BibitemOpen
  \bibfield  {author} {\bibinfo {author} {\bibfnamefont {S.~A.}\ \bibnamefont
  {Ktitorov}}\ and\ \bibinfo {author} {\bibfnamefont {L.}~\bibnamefont
  {Jastrabik}},\ }\href {\doibase 10.1063/1.56292} {\bibfield  {journal}
  {\bibinfo  {journal} {AIP Conference Proceedings}\ }\textbf {\bibinfo
  {volume} {436}},\ \bibinfo {pages} {184} (\bibinfo {year} {1998})},\ \Eprint
  {http://arxiv.org/abs/https://aip.scitation.org/doi/pdf/10.1063/1.56292}
  {https://aip.scitation.org/doi/pdf/10.1063/1.56292} \BibitemShut {NoStop}%
\bibitem [{\citenamefont {Martelli}\ \emph {et~al.}(2018)\citenamefont
  {Martelli}, \citenamefont {Jim\'enez}, \citenamefont {Continentino},
  \citenamefont {Baggio-Saitovitch},\ and\ \citenamefont
  {Behnia}}]{Martelli2018}%
  \BibitemOpen
  \bibfield  {author} {\bibinfo {author} {\bibfnamefont {V.}~\bibnamefont
  {Martelli}}, \bibinfo {author} {\bibfnamefont {J.~L.}\ \bibnamefont
  {Jim\'enez}}, \bibinfo {author} {\bibfnamefont {M.}~\bibnamefont
  {Continentino}}, \bibinfo {author} {\bibfnamefont {E.}~\bibnamefont
  {Baggio-Saitovitch}}, \ and\ \bibinfo {author} {\bibfnamefont
  {K.}~\bibnamefont {Behnia}},\ }\href {\doibase
  10.1103/PhysRevLett.120.125901} {\bibfield  {journal} {\bibinfo  {journal}
  {Phys. Rev. Lett.}\ }\textbf {\bibinfo {volume} {120}},\ \bibinfo {pages}
  {125901} (\bibinfo {year} {2018})}\BibitemShut {NoStop}%
\bibitem [{\citenamefont {Li}\ \emph {et~al.}(2020)\citenamefont {Li},
  \citenamefont {Fauqu\'e}, \citenamefont {Zhu},\ and\ \citenamefont
  {Behnia}}]{Xiaokang2020}%
  \BibitemOpen
  \bibfield  {author} {\bibinfo {author} {\bibfnamefont {X.}~\bibnamefont
  {Li}}, \bibinfo {author} {\bibfnamefont {B.}~\bibnamefont {Fauqu\'e}},
  \bibinfo {author} {\bibfnamefont {Z.}~\bibnamefont {Zhu}}, \ and\ \bibinfo
  {author} {\bibfnamefont {K.}~\bibnamefont {Behnia}},\ }\href {\doibase
  10.1103/PhysRevLett.124.105901} {\bibfield  {journal} {\bibinfo  {journal}
  {Phys. Rev. Lett.}\ }\textbf {\bibinfo {volume} {124}},\ \bibinfo {pages}
  {105901} (\bibinfo {year} {2020})}\BibitemShut {NoStop}%
\bibitem [{\citenamefont {Kustov}\ \emph {et~al.}(2020)\citenamefont {Kustov},
  \citenamefont {Liubimova},\ and\ \citenamefont {Salje}}]{Kustov2020}%
  \BibitemOpen
  \bibfield  {author} {\bibinfo {author} {\bibfnamefont {S.}~\bibnamefont
  {Kustov}}, \bibinfo {author} {\bibfnamefont {I.}~\bibnamefont {Liubimova}}, \
  and\ \bibinfo {author} {\bibfnamefont {E.~K.~H.}\ \bibnamefont {Salje}},\
  }\href {\doibase 10.1103/PhysRevLett.124.016801} {\bibfield  {journal}
  {\bibinfo  {journal} {Phys. Rev. Lett.}\ }\textbf {\bibinfo {volume} {124}},\
  \bibinfo {pages} {016801} (\bibinfo {year} {2020})}\BibitemShut {NoStop}%
\bibitem [{\citenamefont {Weber}(2021)}]{WEBER}%
  \BibitemOpen
  \bibfield  {author} {\bibinfo {author} {\bibfnamefont {T.}~\bibnamefont
  {Weber}},\ }\href {\doibase https://doi.org/10.1016/j.softx.2021.100667}
  {\bibfield  {journal} {\bibinfo  {journal} {SoftwareX}\ }\textbf {\bibinfo
  {volume} {14}},\ \bibinfo {pages} {100667} (\bibinfo {year}
  {2021})}\BibitemShut {NoStop}%
\bibitem [{\citenamefont {Yamanaka}\ \emph {et~al.}(2000)\citenamefont
  {Yamanaka}, \citenamefont {Kataoka}, \citenamefont {Inaba}, \citenamefont
  {Inoue}, \citenamefont {Hehlen},\ and\ \citenamefont
  {Courtens}}]{Yamanaka2000}%
  \BibitemOpen
  \bibfield  {author} {\bibinfo {author} {\bibfnamefont {A.}~\bibnamefont
  {Yamanaka}}, \bibinfo {author} {\bibfnamefont {M.}~\bibnamefont {Kataoka}},
  \bibinfo {author} {\bibfnamefont {Y.}~\bibnamefont {Inaba}}, \bibinfo
  {author} {\bibfnamefont {K.}~\bibnamefont {Inoue}}, \bibinfo {author}
  {\bibfnamefont {B.}~\bibnamefont {Hehlen}}, \ and\ \bibinfo {author}
  {\bibfnamefont {E.}~\bibnamefont {Courtens}},\ }\href {\doibase
  10.1209/epl/i2000-00325-6} {\bibfield  {journal} {\bibinfo  {journal}
  {Europhysics Letters ({EPL})}\ }\textbf {\bibinfo {volume} {50}},\ \bibinfo
  {pages} {688} (\bibinfo {year} {2000})}\BibitemShut {NoStop}%
\end{thebibliography}%

\end{document}